%Paper: hep-ph/9303212
%From: christoc@neutron.tp2.ruhr-uni-bochum.de (Christo Christov)
%Date: Wed, 3 Mar 1993 14:05:49 +0100

% file: TXSsymb.tex  TeXsis   version 2.10
%======================================================================
% EXTENDED MATH SYMBOLS FOR PHYSICS
%
% This file defines extensions to the symbols in Plain TeX that are
% generally useful for physics papers, espacially for high energy physics.
% See the comments for each definition.
%
%-----------------------------------------------------------------------
\message{Extended math symbols.}

%-----------------------------------------------------------------------
% Raise \chi and \zeta so they do not have descenders. This looks
% better in formulae.
\ifx\oldzeta\undefined    % hasn't been done yet, so
  \let\oldzeta=\zeta    % save old definiton
  \def\zzeta{{\raise 2pt\hbox{$\oldzeta$}}} % make new definition
  \let\zeta=\zzeta    % and attatch it
\fi

\ifx\oldchi\undefined    % hasn't been done yet, so
  \let\oldchi=\chi    % save old definiton
  \def\cchi{{\raise 2pt\hbox{$\oldchi$}}} % make new definition
  \let\chi=\cchi    % and attatch it
\fi

%-----------------------------------------------------------------------
% Various special symbols.

% Gradient, etc.
 % box
    % gradient
    % synonym for \partial

% Fractions.
\def\frac#1#2{{#1 \over #2}}

\def\half{\ifinner {\scriptstyle {1 \over 2}}
   \else {1 \over 2} \fi}

% Bras and kets
  % \bra{stuff} gives <stuff|
  % \ket{stuff} gives |stuff>

% \simge and \simle make the "greater than about" and the "less
% than about" symbols with spacing as relations.
\def\simge{\rlap{\raise 2pt \hbox{$>$}}{\lower 2pt \hbox{$\sim$}}}
\def\simle{\rlap{\raise 2pt \hbox{$<$}}{\lower 2pt \hbox{$\sim$}}}

% \parenbar puts a bar in small parentheses over a character to
% indicate an optional antiparticle. \nunubar and \ppbar are special
% cases.
     % right-hand spacing

% \buildchar makes a compound symbol, placing #2 above #1 and #3
% below it with \limits. \overcirc is a special case.
     % cancel space, \null

% \slashchar puts a slash through a character to represent contraction
% with Dirac matrices. Use \not instead for negation of relations, and use
% \hbar for hbar.
\def\slashchar#1{\setbox0=\hbox{$#1$}  % set a box for #1
   \dimen0=\wd0     % and get its size
   \setbox1=\hbox{/} \dimen1=\wd1  % get size of /
   \ifdim\dimen0>\dimen1   % #1 is bigger
      \rlap{\hbox to \dimen0{\hfil/\hfil}} % so center / in box
      #1     % and print #1
   \else     % / is bigger
      \rlap{\hbox to \dimen1{\hfil$#1$\hfil}} % so center #1
      /      % and print /
   \fi}      %

%-----------------------------------------------------------------------
% Various expressions in Roman type.

% Functions -- all defined like \sin, etc. in Plain TeX:
 % Re for real part
 % Im for imaginary part

 % tr for trace
 % Tr for functional trace
 % Det for functional determinant

 % mod for modulo
 % wrt for with respect to

% Abbreviations.
   % 10^12 electron volts
   % 10^9  electron volts
   % 10^6  electron volts
   % 10^3  electron volts
   % 1     electron volt

   % 10^-27 cm^2
   % 10^-30 cm^2
   % 10^-33 cm^2
   % 10^-36 cm^2

% >>> EOF TXSsymb.tex <<<
%>>>>>>>>>>>>>>>>>>techmac.tex<<<<<<<<<<<<<<<<<<<<<<<<<<<<<<<<<<<<<<
  \def\za{\alpha}          
  \def\zb{\beta}           
            
  \def\ze{\epsilon}         
              
  \def\zg{\gamma}

  \def\zl{\lambda}         \def\zL{\Lambda}
  \def\zm{\mu}

  \def\zp{\pi}             
              \def\zQ{\Psi}
              
  \def\zs{\sigma}          
  \def\zt{\tau}
  
  \def\zw{\omega}

%>>>>>>>>>>>>>>>>>>additional definitions<<<<<<<<<<<<<<<<<<<<<<<<<<<

\def\fpi{f_\pi}
\def\mpi{m_\pi}

\def\vpi{\vec\pi}
\def\vtau{\vec\tau}

\def\rnm{\varrho_{nm}\,}

%***  this is file PHYSIKP as of 10.2.89  ***
\def\wlog#1{} % suppress allocation messages
\catcode`\@=11

\outer\def\rthnum#1{\topright{RUB-TPII-#1}}

\def\wlog{\immediate\write\m@ne} % restore PLAIN's definition
\catcode`\@=12 % @ signs are no longer letters

%***  additional physics journals  ***

\def\AP{\journal{Ann.\ \Phys}}

\def\NC{\journal{Nuov.\ Cim.\ }}

\def\ZPC{\journalp{Z.\ \Phys}C}

\def\APPB{\journalp{Acta \ \Phys \ Polonica}B}

%***  end of file PHYSIKP  ***
\input phys
% NJL MEDIUM                                      March  1993
%=======================================================================
%
\english
\titlepage
\chapters %for normal paper
\equchap
\equfull
\figpage
\tabpage
\refpage
\overfullrule=0pt
%
%******************* references **************************************
%
\RF\Egu76{T.Eguchi, \PRD14(1976)2755*}
\RF\Ebert83{D.Ebert and M.K.Volkov, \ZPC16(1983)205*}
\RF\Ebert86{D.Ebert and H.Reinhardt, \NPA271(1986)188*}
\RF\Bernard86{V.Bernard, \PRD34(1986)1601*}
\RF\Bernard84{V.Bernard, R.Brockmann, M.Schaden, W.Weise and E.Werner,
\NPA412(1984)349*}
\RF\Provid87{J.da Providencia,M.C.Ruivo and C.A.de Sousa, \PRD36(1987)1882*}
\RF\Diakonov88{D.I.Diakonov, V.Yu.Petrov and P.V.Pobylitsa, \NPB306(1988)809*}
\RF\Ripka84{S.Kahana and G.Ripka, \NPA429(1984)462*}
\RF\Ripka91{G.Ripka and R.M.Galain, Lecture at the XXX Cracow School of
Theoretical Physics, June 2-12, 1990, Zkopane, Poland, {\APPB22(1991)187*}
; M.Jaminon, R.M.Galain, G.Ripka and P.Stassart, \NPA537(1992)418*}
\RF\Shifman79{M.A.Shifman, A.J.Vainstein and V.I.Zakharov,
{\NPB147(1979)385*}; \NPB163(1980)43*}
\RF\Nam61{Y.Nambu and G.Jona-Lasinio, \PR122(1961)354*}
\RF\Schwinger51{J.Schwinger, \PR82(1951)664*}
\RF\GellMann60{M.Gell-Mann and M.L\'evy, \NC16(1960)705*}
\RF\Chr90{Chr.V.Christov, E.Ruiz Arriola and K.Goeke,
{\NPA510(1990)1990*}; \PLB225(1989)22*}
\RF\Reinhardt88{H.Reinhardt and R.W\"unsch, {\PLB215(1988)577*};
\PLB230(1989)93*}
\RF\Prasz90{M.Praszalovicz, \PRB42(1990))216*}
\RF\Alkofer90{R.Alkofer, \PLB236(1990)461*}
\RF\TMeissner88{T.Meissner,E.Ruiz Arriola, F.Gr\"ummer, K.Goeke,
H.Mavromatis, \PLB214(1988)312*}
\RF\Goeke91{K.Goeke, A.G\'orski, F.Gr\"ummer, Th.Mei\ss ner, H. Reinhardt
and R. W\"u nsch, {\PLB256(1991)321*};
A.G\'orski, F.Gr\"ummer and K.Goeke, \PLB278(1992)24*}
\RF\Wakamatsu91{M.Wakamatsu and H. Yoshiki, \NPA524(1991)561*}
\RF\TMeissner89{ T.Meissner, F.Gr\"ummer and K.Goeke, {\PLB227(1989)296*};
{\AP202(1990)297*}; T.Meissner and K.Goeke, \NPA524(1991)719*}
\RF\UMeissner89{U.-G.Meissner, {\PRL62(1989)1012*};\PLB220(1989)1*}
\RF\Bernard87{V.Bernard, Ulf-G.Meissner and I.Zahed, {\PRD36(1987)819*};
{\PRL59(1987)966*};
V.Bernard and Ulf-G.Meissner, \NPA489(1989)647*}
\RF\Jaminon89{M.Jaminon, G.Ripka and P.Stassart, {\NPA504(1989)733*};
M.Jaminon, R.Mendez Galain, G.Ripka and P.Stassart, \NPA537(1992)418*}
\RF\Walecka74{J.D.Walecka, {\AP83(1974)491*}; B.D.Serot and
J.D.Walecka, Adv. in Nucl. Phys. v.16, eds. J.W.Negele and E.Vogt
(Plenum Press, New York, 1986)}
\RF\Mahaux87{C.Mahaux and R.Sartor, \NPA475(1987)247*}
\RF\Fukugita88{for reviews, see M.Fukugita, {\NPB
(Proc.Suppl.)4(1988)105*}; {\NPB (Proc.Suppl.)9(1989)291*}; A.Ukawa,
{\NPB (Proc.Suppl.)10A(1989)66*}; \NPA498(1989)227c*}
\RF\Lutz92{M.Lutz, S.Klimt and W.Weise, \NPA542(1992)531*}
%
%************************ figure captions ********************************
%
\FIG\Figr1{Constituent quark mass $M^*$ as a function of the medium baryon
density. At vanishing density it is fixed to $M=420$ MeV. The solid line
represents the present calculations with a finite quasi-discrete basis
whereas the dashed one - the calculations with plane-wave basis.}
\FIG\Figr2{Soliton mass as a function of the medium baryon density for two
different values of the constituent quark mass is compared to the mass of
the soliton (dash-dotted line) of the medium-modified $\zs$-model.}
\FIG\Figr3{Soliton mass and the contributions coming from the
valence quarks and from the polarized continua as a function of the medium
baryon density for the constituent mass $M=420$ MeV. The separated
contributions from the polarization of the Fermi and Dirac sea are shown as
well.}
\FIG\Figr4{Density distribution of the soliton in medium for
three different medium baryon densities. The constituent mass is $M=420$
MeV.}
\FIG\Figr5{Soliton mean square radius as a function of the medium baryon
density for two different values of the constituent quark mass.}
%
%**************************** paper **************************************
%
\rthnum{26/92}

\title{B=1 Soliton of the Nambu - Jona-Lasinio model in medium}

\author{Chr.V.Christov{\rm\footnote{${}^{\dagger}$}{Permanent
address: Institute for Nuclear Research and Nuclear Energy, Sofia 1784,
Bulgaria}} and K.Goeke}

\address{Institut f\"ur Theoretische Physik II, Ruhr-Universit\"at
Bochum, \nl D-4630 Bochum}

\vskip 1cm

\abstract{The solitonic sector of
the Nambu - Jona-Lasinio model with baryon number one is solved in
the presence of an external medium. The calculations fully include the
polarization of both the Dirac sea and the medium as well as the Pauli
blocking effect. We found that with an increasing density the medium
destabilizes the soliton. At finite medium density the soliton mass gets
reduced whereas the mean  square baryon radius shows an increase - a swelling
of the soliton. At some critical density of about two times nuclear
matter density there is no localized solution - the soliton disappears.}

\endpage

\chap{Introduction}

The chiral symmetry is believed to be a dominant
feature of the low-energy sector of QCD and in particular,
in the structure of the low-lying hadrons.
In the present absence of a solution to the non-perturbative
QCD, different effective models have
been employed to study the meson and baryon structure. Among them
the Nambu - Jona-Lasinio model\quref{\Nam61}) seems to
play an essential role. From one hand the model accounts for the the
chiral symmetry breaking - describes a structured vacuum
and the mesons as elementary excitations of the non-vanishing  $<\bar qq>$
condensate and from the other - it incorporates the general accepted
phenomenological picture of the baryon as a bound state of
$N_c$ valence quarks coupled to the quarks of the polarized Dirac sea. Quite
encouraging, the model shows a significant success in describing the meson
sector\quref{\Ebert83\use{\Bernard84\Bernard86\Ebert86}-\Provid87}) as
well as the nucleon as a baryon number one
soliton\quref{\Diakonov88\use{\Reinhardt88\TMeissner88\TMeissner89\Prasz90
\Alkofer90\Wakamatsu91}-\Goeke91}).

As it is suggested by the lattice gauge QCD
calculations\quref{\Fukugita88}), at some finite
temperature and/or density one generally expects a restoration of the chiral
symmetry and a deconfinement, and hence a change of the hadron structure.
Following this idea to consider the restoration
of the chiral symmetry in a medium as a relevant mechanism for a scale
change which modifies the hadron structure, the NJL model has been also
applied to study medium modifications of the meson
properties\quref{\Bernard87\use{\Jaminon89\Chr90}-\Lutz92}). Indeed,
adding a medium (partially occupied positive continuum)
the model allows for the (partial) restoration of the chiral symmetry at
finite density and/or temperature and thus, the model seems to provide an
appropriate working scheme to study modifications of the meson
properties in medium. Therefore, it is reasonable to apply the NJL model to
study the structure of the more complicated objects like baryons in
medium since the model allows one to treat exactly the
polarizations of both the medium and the Dirac sea as well as
the Pauli blocking effect. It is the purpose of present work to
solve directly the B=1 soliton sector of the NJL model in a
medium of finite density.

\chap{NJL model with a chemical potential}

We start with the lagrangean of the NJL  model with scalar and
pseudoscalar quark-quark couplings\quref{\Nam61}):
$$
{\cal L}=\overline\Psi i\zg^\zm\partial_\zm\zQ-m_0\overline\zQ\zQ
+\frac G2\lbrack(\overline\zQ\zQ)^2+(\overline\zQ\zg_5\vec\zt\zQ)^2
\rbrack.\EQN\Eq1
$$
Here $\Psi$ describes a quark field of $SU(2)$-flavour ($u$ and $d$
quarks), $N_c=3$ colours and average current mass $m_0=(m_u+m_d)/2$.
Introducing auxiliary sigma and pion fields by
$\zs=-g\overline\zQ\zQ/\zl^2$ and
${\vec\zp}=-g\overline\zQ i\zg_5\vec\zt\zQ/\zl^2$ lagrangean \queq{\Eq1}
can be rewritten in the form:
$$
{\cal L}=\overline\Psi
[i\zg^\zm\partial_\zm-g(\zs+i\zg_5\vpi.\vtau)]\Psi
+\frac {\zl^2}2(\zs^2+\vpi^2)-\frac {m_0\zl^2}g \zs, \EQN\Eq1a
$$
Assuming the boson fields to
be classical (zero boson and one fermion loop approximation), one can
use path integral techniques in euclidean metrics to evaluate the
corresponding effective action $S_{eff}$ as a sum of a quark part
$$
S_{eff}^q=Sp\log\left(\frac \partial{\partial t}+h-\mu\right) \EQN\Eq3
$$
and a meson one
$$
S_{eff}^m=\int d^4x\left\{\frac
{\zl^2}2(\zs^2+\vpi^2)-\mpi^2\fpi\zs\right\}.
\EQN\Eq2
$$
Here $\zm$ is the chemical potential and $h$ is the single particle
Hamiltonian
$$
h=\frac {\vec\za.\vec\nabla}i+\zb g(\zs+i\zg_5\vpi.\vtau), \EQN\Eq4
$$
$Sp$ means the functional and matrix (spin, isospin and colour)
trace. In the meson part, using the PCAC, the current mass $m_0$ is
eliminated
in favor of the pion mass  $\mpi$ and the pion decay constant $\fpi$.
The coupling constant $G$ is related to $\zl^2$ by $G=g^2/\zl^2$
where an additional coupling constant $g$ is introduced for convenience.
Actually, in order to fix our parameter by reproducing the pion decay
constant $\fpi$,
$g$ will be fixed later on to the physical coupling constant in order
to identify the introduced auxiliary fields with the physical pion and sigma
mesons.

Due to the local four-fermion interaction the lagrangean
\queq{\Eq1} is not renormalizable and a regularization procedure with an
appropriate cutoff $\Lambda$ is needed to make the effective action
finite. Actually, only a part of the effective action,
$S_{eff}^q(\zm=0)$, coming from the Dirac sea is divergent. Using the
proper-time regulator\quref{\Schwinger51}):
$$
\log \za \Rightarrow -\int_{\zL^{-2}}^\infty \frac {d\zt}{\zt}
e^{-{\zt\za}}, \EQN\Eq4a
$$
we regularize it. The rest is the contribution coming from the valence and
medium quarks and it is finite. One can express the energy of the system
(see for more details refs.\quref{\TMeissner89,\Chr90}) as a sum of a quark
part
$$
E^q=\eta_{val}\ze_{val}+\sum\limits_{M\le\ze_\za\le\zm}\ze_\za+\frac
{N_c}2\sum\limits_\za R_{3/2}(\ze_\za,\zL) \EQN\Eq5
$$
with the proper-time regularization function
$$
R_a(\ze,\zL)=\frac 1{\sqrt{4\zp}}\int\limits_{1/\zL^2}^\infty \frac
{d\zt}{\zt^a}\exp({-\ze^2\zt}), \EQN\Eq6
$$
and the meson part which up to a trivial factor
is given by eq.\queq{\Eq2}. Here $M$ is the constituent quark mass, the
energies $\ze_\za$ are the eigenvalues of
the hamiltonian \queq{\Eq4}
$$
h\phi_\za=\ze_\za\phi_\za, \EQN\Eq6a
$$
and
$$
\eqalign{&\eta_{val}=1 \quad \hbox{for} \quad \ze_{val}>0 \cr
&\eta_{val}=0 \quad \hbox{for} \quad \ze_{val}< 0 \cr}. \EQN\Eq34
$$
The baryon density $\varrho(r)$ can be split in valence, medium and sea
parts:
$$
\varrho_{val}({\bf r})=\eta_{val}\phi_{val}^\dagger({\bf r})\phi_{val}({\bf
r}), \EQADV\Eq7\SUBEQNBEGIN\Eq7a
$$
$$
\varrho_{med}({\bf
r})=\sum\limits_{M\le\ze_\za\le\zm}\phi_\za^\dagger({\bf
r})\phi_\za({\bf r}), \SUBEQN\Eq7b
$$
$$
\varrho_{sea}({\bf
r})=\frac 12\sum\limits_\za\phi_\za^\dagger({\bf
r})\phi_\za({\bf r})sign(-\ze_\za), \SUBEQN\Eq7c
$$

Using the spectral representation of the energies $\ze_\za$ the equations of
motions for the meson fields
$$
\frac {\delta S_{eff}}{\delta \zs}=0 \qquad
\hbox{and} \qquad \frac {\delta S_{eff}}{\delta
\vpi}=0 \EQN\Eq8
$$
can be written as
$$
\eqalign{\zs=\frac g{\zl^2}N_c\Bigl\{ \sum\limits_\za
\overline\phi_{\za}\phi_{\za}
R_{1/2}(\ze_{\za},\zL)&-\eta_{val}\overline\phi_{val}\phi_{val}\cr
&-\sum\limits_{M\le\ze_\za\le\zm}\overline\phi_{\za}\phi_{\za}\Bigr\}
+\frac{\fpi\mpi^2}{\zl^2} \cr} \EQADV\Eq9\SUBEQNBEGIN\Eq9a
$$
$$
\eqalign{\vpi=\frac g{\zl^2}N_c\Bigl\{
\sum\limits_\za
\overline\phi_{\za} i \zg_5\vec\zt\phi_{\za}
R_{1/2}(\ze_{\za},\zL)&-\eta_{val}\overline\phi_{val} i
\zg_5\vec\zt\phi_{val} \cr
&-\sum\limits_{M\le\ze_\za\le\zm}\overline\phi_{\za} i
  \zg_5\vec\zt\phi_{\za}\Bigr\}.\cr} \SUBEQN\Eq9b
$$

\chap{Fixing of model parameters}

Apart from the current mass $m_0$ (eliminated using PCAC) the NJL model
contains also as parameters the coupling constant $G$ (or $\zl^2$ in
eq.\queq{\Eq1a}) and cutoff $\zL$. We fix these parameters in the meson
(translationally invariant field configuration) sector of the vacuum.

First, in order to express the coupling constant $\zl^2$ as a function of the
constituent mass $M=g<\zs>_v$, where $<\zs>_v$ is the expectation value of
the $\zs$ field in vacuum, we use the equation of motion of the sigma field
eq.\queq{\Eq9\Eq9a} in vacuum (the gap equation):
$$
\zl^2=N_c g^2 \frac 1{\zp^{3/2}} R_{2}(M,\zL)+\mpi^2 \EQN\Eq11
$$
The equation of motion of the pioneq field eq.\queq{\Eq9\Eq9b} gives a
trivial solution in vacuum $<\vpi>_v=0$.

Second, the cut-off $\zL$ is fixed by reproducing the experimental values
of the pion decay constant $\fpi$ and the pion mass $\mpi$ in the meson
sector of the vacuum. This procedure originates from Eguchi\quref{\Egu76})
and is described in detail in ref.\quref{\TMeissner89}). This fixes the
expectation value of the $\zs$ in vacuum
$$
<\zs>_v=\fpi \EQN\Eq11a
$$
and leads to the following relation between the constituent mass $M$ and
the cutoff $\zL$:
$$
N_c g^2 \frac 1{2\zp^{3/2}} R_1(M,\zL)=1. \EQN\Eq12
$$
It is easy to check that this condition insures (in soft pion limit
$\mpi\approx 0$)
that the auxiliary fields $\zs$ and $\vec\zp$ are the physical meson fields.
Thus, in this scheme the vacuum value of the constituent mass $M=g\fpi$ is
the only free parameter. In principle, the empirical
values\quref{\Shifman79}) of the quark condensate and the quark current mass
can be used to fix $M$ but actually they still leave a broad range for $M$.

At finite chemical potential we use unchanged the values of
$\zl^2$ and $\zL$. We also keep $g$ fixed as a particular scale. It should
be noted, however, that the solitonic solution does not depend on the
particular way in which the auxiliary boson fields are scaled.

\chap{Numerical procedure}

Similar to other works\quref{\Reinhardt88,\TMeissner89,
\Alkofer90,\Wakamatsu91}) we also use the finite
quasi-discrete basis and numerical method of Ripka and
Kahana\quref{\Ripka84}) for solving the eigenvalue problem (eq.\queq{\Eq6a})
by putting the system in a spherical box of a large radius $D$. The basis is
made discrete by imposing a boundary condition at $D$. It is also made
finite by restricting the momenta of the basis state to be smaller
than a numerical cutoff $K_{max}$. Obviously both quantities have no any
physical meaning and the results should not depend on them. The typical
values, which we use, are $D\approx 20/M$ and $K_{max}\approx 7M$. In
addition all checks\quref{\TMeissner89,\Wakamatsu91}) concerning the
numerical stability of the solution with respect to varying the box size
and the choice of the numerical cut-off have been done such that the actual
calculation is completely under control.

\chap{Meson sector}

As a first step we look for a translationally invariant solution (meson
sector) $\partial_\zm\zs=\partial_\zm\vec\zp=0$ at
vanishing as well as at finite chemical potential using
the finite quasi-discrete basis\quref{\Ripka84}). Actually, for the
solitonic solution later we need
the translationally invariant meson field configuration, which makes the
effective action $S_{eff}$ stationary.
To this end we solve the equations of motion eqs.\queq{\Eq9\Eq9a,\Eq9\Eq9b}
together with Dirac equation \queq{\Eq6a} in the case of constant fields.
For the pion
field we get the trivial solution $<\vec\zp>_\zm=0$ (expectation
value at finite chemical potential) whereas the equation
of motion of the sigma field gives the medium-modified value of the
constituent quark mass\footnote{$^\ddagger$}{the asterisk means
medium-modified value} $M^*=g <\zs>_\zm$ :
$$
M^*=N_c \frac {g^2}{\zl^2}\left\{ \sum\limits_\za
\overline\phi_{\za}\phi_{\za} R_{1/2}(\ze_{\za},\zL)-
\sum\limits_{M^*\le\ze_\za\le\zm}\overline\phi_{\za}\phi_{\za}\right\}
+\frac{M\mpi^2}{\zl^2}. \EQN\Eq13
$$

For a particular value of the chemical potential the baryon density of the
medium is given by eq.\queq{\Eq7\Eq7b}. In this case the contribution of the
Dirac sea to the baryon density is exactly zero. It should be also noted
that due to the boundary conditions on the basis functions, the medium
density deviates from a constant value  in the vicinity of the box surface.
In order to diminish this finite size effects we always pay
attention to have a box size large enough with respect to the size of the
soliton. To be particular, at fixed medium density the results show
within 5 \% no dependence on both the box radius and the corresponding
number of the
quarks in the box. The stability of our translationally invariant solution
with respect to finite size effects is illustrated in \qufig{\Figr1}
where the constituent mass $M^*$, calculated in the Kahana-Ripka basis as a
function of the baryon density, is compared with the results\quref{\Chr90})
obtained using a plane wave basis (no finite size effects). As can be seen
that up to one and a half nuclear matter density ($\rnm$) the two solutions
almost coincide and start to deviate slightly at higher density values.
It shows that the finite basis provides a satisfactorily good
description of the positive energy continuum (medium). Needless to say
that the reduction of the constituent mass, found at finite density,
indicates a trend towards a restoration of the chiral symmetry in
medium.

\chap{B=1 solitonic (nucleon) sector}

As a next step we solve the solitonic sector in medium. We look for a
localized bound solution of three valence quarks interacting with the Fermi
and Dirac sea quarks all of a constituent mass $M^*$. We use a numerical
self-consistent iterative procedure\quref{\TMeissner89}) based on a method
proposed by Ripka and Kahana\quref{\Ripka84}). The procedure consists in
solving in an iterative way the Dirac equation \queq{\Eq4} together with the
equations of motion of the meson fields - eqs.\queq{\Eq9\Eq9a,\Eq9\Eq9b}.
The meson fields are assumed to be in a hedgehog form
$$
\zs({\bf r})=\zs(r) \qquad \hbox{and} \qquad \vec\zp({\bf r})={\bf \hat
r}\zp(r)
 \EQN\Eq14
$$
and are restricted on the chiral circle
$$
\zs^2+{\vpi}^2={M^*}^2/g^2. \EQN\Eq15
$$
The techniques for the numerical procedure are well
known\quref{\Reinhardt88,\TMeissner89,\Alkofer90}) for the case
of vanishing chemical potential where there is no contribution coming from
the positive continuum. They can be easily adopted to the case of finite
chemical potential. As can be seen the latter leads to an additional source
(Fermi sea quarks) for the meson fields in eqs.\queq{\Eq9\Eq9a,\Eq9\Eq9b}.

Because of the presence of the soliton both the
Fermi sea and the Dirac sea get polarized. Subtracting the energy of the
unperturbed Fermi and Dirac sea (translationally invariant
solution) the B=1 soliton energy (effective soliton mass) is given
by the sum of the energy of the valence quarks and
the contributions due to the polarization of both continua:
$$
\eqalign{
E=\eta_{val}\ze_{val}+\sum\limits_{M\le\ze_\za\le\zm}(\ze_\za-\ze_\za^0)
&+\frac {N_c}2\sum\limits_\za\left\{R_{3/2}(\ze_\za,\zL)
-R_{3/2}(\ze_\za^0,\zL) \right\} \cr
&+\fpi\mpi2\int d^3r\{M*/g-\zs(r)\}. \cr} \EQN\Eq16
$$
The energies $\ze_\za^0$ correspond to the translationally invariant
solution.

The density dependence of the B=1 soliton mass calculated for two different
values of the constituent mass $M=420$ and $M=465$ MeV is shown in
\qufig{\Figr2}. The two curves almost coincide and both show a strong
reduction of the soliton mass at finite density. Hence one should
conclude that at least for the
constituent mass between 400 and 470 MeV this medium effect seems to be
practically independent on the particular value of $M$ (vacuum value).
For comparison we also present the mass of the soliton of the
medium-modified $\zs$-model
lagrangean\quref{\Chr90}). As can be seen the latter shows a
weaker density dependence. For instance, at $\rnm$ it was around 20 \%
whereas the present calculations give about 30\%. This number is close
to the estimate coming from the $\zs-\zw$ model\quref{\Walecka74})
whereas the many-body nuclear theory\quref{\Mahaux87}) gives a smaller
number similar to results of the $\zs$-model. It should be noted, however,
that these calculations\quref{\Chr90}) suffer from the fact that the
Fermi sea is not included and the Pauli blocking is not taken
into account.

At about two times nuclear matter density we found
no solitonic solution. It means that at some critical density the inclusion
of the medium destroys the condition for the existence of a localized
solution in the model. Similar effect is found\quref{\Chr90,\UMeissner89})
in other effective models like the Skyrme model with vector mesons and the
Gell-Mann - L\'evy $\zs$-model. It is rather tempting to interpret this
effect as a an indication of a delocalization of the nucleon in a dense
medium but such a conclusion may go out of the scope of the NJL model
itself because of the lack of confinement.

It is also interesting to look at the separate contributions to the
soliton energy coming from the valence quarks as well as from the Fermi
and Dirac sea (see \qufig{\Figr3}). At constituent mass $M=420$ MeV in
the vacuum as well as in medium about a half of the soliton energy
comes from the polarized Fermi and Dirac sea. With
increasing medium density both the valence and the summed sea (Fermi+Dirac)
contribution show a similar reduction. It should be noted, however, that
the contribution of the Fermi sea alone is negative
(opposite sign) and shows a different behaviour. Apart from
the sign the contribution of the polarized Dirac sea shows a similar density
dependence. It means that with increasing density the polarized Fermi sea
trends towards
a destabilization of the soliton whereas the polarized Dirac sea stabilizes
it. As can be seen the contribution of the Dirac sea takes over and the
summed sea contribution, which is a result of a cancellation of two large
numbers, is positive. Apparently, at least in the present model
calculations the contribution, coming from the polarized Fermi sea, plays an
important role and can not be neglected. Close to
the critical density the contributions of the polarized Fermi and Dirac
compensate each other and the summed sea contribution tends to vanish
whereas the valence energy saturates. Since the localized solution exists
because of the interplay between the valence and the sea contribution the
vanishing of the latter leads to a destabilization of the soliton.
This picture can be understood in the context of the partial
restoration of the chiral symmetry, i.e. reduction of the chiral order
parameter and of the constituent mass. The latter means a decrease of
the gap of the Dirac spectrum. It leads to a reduction of the
valence quark energy. We observe also that the Fermi and the Dirac sea are
getting polarized at the same rate. Thus, the
summed sea contribution vanishes and the soliton gets destabilized.

Similar to the energy the baryon density distribution of the soliton
$\varrho_{sol}(r)$ can be split into valence and sea parts:
$$
\eqalign{
\varrho_{sol}(r)=\eta_{val}\phi_{val}^\dagger(r)\phi_{val}(r)&+
\sum\limits_{M^*\le\ze_\za\le\zm}\{\phi_\za^\dagger(r)\phi_\za(r)
-{\phi_\za^0}^\dagger(r)\phi_\za^0(r)\} \cr
&+\frac 12\sum\limits_\za\phi_\za^\dagger(r)\phi_\za(r)sign(-\ze_\za).\cr}
\EQN\Eq17
$$
The wave functions $\phi_\za^0$ correspond to the translationally invariant
solution.

It is easy to check thet the baryon number of the solitonic solution is one.
The effect of the medium on the spatial distribution
of the soliton is illustrated in \qufig{\Figr4} where the baryon density of
the soliton is plotted for different values of the medium density. It is
clear that at finite density the soliton gets more extended which supports
the idea of swelling of the nucleon in medium. Close to two times nuclear
matter density the solution shows a clear delocalization: the distribution
is rather close to the uniform one.

As a measure of the spatial extension of the soliton in medium we also
calculated the mean square radius of the soliton:
$$
<r^2>_{sol}=\int d^3r\, r^2\varrho_{sol}({\bf r}). \EQN\Eq18
$$
The results
are depicted in \qufig{\Figr5}. As it is expected from the soliton density
distributions in medium (shown in \qufig{\Figr4}) the radius
increases with increasing medium density. At densities close to the
critical one it seems to diverge. Similar to the mass
the relative change of the radius is larger than the
estimates\quref{\Chr90})
in $\zs$-model: at $\rnm$ the rms radius is increased by 30\%
whereas the solitonic sector of the $\zs$-model gives an enhancement of
about 20\%.

\chap{Summary}

We solved self-consistently the baryon number one solitonic sector of NJL
model in order to investigate the structure of the solitonic solution with
baryon number $B=1$ immersed in a quark medium of finite density.
medium in a self-consistent way. The polarization of both continua, namely
the Fermi and the Dirac sea, as well as the Pauli blocking effect are
treated on the same footing. We found that the polarization of the medium
plays a role as important as the polarization of the Dirac sea and
with increasing density the medium polarization destabilizes
the soliton. The B=1 solitonic solution obtained gets more extended
in medium (swelling). The mass gets reduced whereas the mean square radius
increases. The relative changes
of both quantities are larger than the estimates\quref{\Chr90}) obtained
from the soliton sector of
the medium-modified Gell-Mann - L\'evy $\zs$-model.
At increasing medium density the soliton show a clear trend to a
delocalization and at some critical value, less than two times nuclear
matter density, it disappears.

{\it The work has been supported by the Bundesministerium f\"ur
Forschung und Technologie, Bonn (Int. B\"uro), by the KFA J\"ulich (COSY --
Project) and the Deutsche Forschungsgemeinschaft.}

\refout
\figout
\end